\documentclass[a4paper]{article}

\usepackage{fullpage,url,amsmath,amsfonts,amssymb,tedmath}

\newcommand{\di}{\displaystyle}
\newcommand{\al}{\alpha}
\newcommand{\om}{\omega}
\newcommand{\be}{\beta}
\newcommand{\de}{\delta}
\newcommand{\ga}{\gamma}
\newcommand{\F}{\mathbb{F}}
\newcommand{\tri}{\triangle}
\newcommand{\ti}{\times}

\title{Self-dual, dual-containing and related quantum
  codes from group rings.}
\author{Ted Hurley}
\date{}
\begin{document}
\maketitle
\begin{abstract}
Classes  of
self-dual codes and  
dual-containing codes are constructed. The  codes are obtained 
within group rings and, using an isomorphism
between group rings and matrices, equivalent codes are obtained in matrix
form. Distances and other properties are derived by working within
the group ring. Quantum codes are constructed from the  dual-containing
codes. 
\end{abstract}

\section*{Introduction}

Classes of self-dual codes and dual-containing codes 
 are constructed using the module methods of 
group ring zero-divisor and unit-derived codes
 as defined in 
 \cite{hur1}. 

Distances and other properties are determined 
algebraically working within the group ring.  
Generator and check matrices for equivalent matrix codes are
 immediately derivable from the group ring constructions using an
 isomorphism between a group ring and a ring of matrices, see \cite{hur2}.  

The group ring methods of \cite{hur1} 
 expand the
range of codes available and  open
up  new series of codes. The methods can be used to derive properties as
well as to construct codes with particular properties. 
Properties of codes, such as being self-dual or low density, have in
many instances 
easy formulations as properties
in group rings and these ideas can be exploited to construct codes with
desired properties. Elsewhere  these group ring methods are used to
 algebraically construct  LDPC
(Low Density Parity Check) codes. In \cite{hur4} a variation in the
 method of unit-derived codes, using group rings over rings which are
 themselves group rings,  is  used to construct  
 classes of convolutional codes and to derive properties thereof. 

Here the group ring   methods are used  to  construct self-dual and 
dual-containing codes.  Quantum codes can  then be constructed from these
 dual-containing codes by the method of \cite{calderbank}. 

General  methods to construct self-dual and dual-containing codes from
modules in group rings  are derived. The general methods  are then
specialised to  particular  group rings to
 derive the classes.  The codes derived are a selection of what
 can be achieved from the general method.

The codes
 obtained  are  not   cyclic, quasi-cyclic  nor
 shortened cyclic codes.  
Specific groups of the group rings used  to
construct the classes of codes derived here include  
 direct products of cyclic groups, 
dihedral groups and generalised dihedral
 groups.


\section{Statement of results} 

\subsection{Initial constructions}\label{sec:construct}
The following classes of binary self-dual codes are initially
constructed.
\begin{itemize}
\item  Class 1: $(2\times 4^m, 4^m, 2\times
 3^{\frac{m}{2}})$ codes for  $m$ even;  
$(2\times 4^m, 4^m, 4\times 3^{\frac{m-1}{2}})$ codes for $m$ odd.
\item Class 2: $(2\times 6^m, 6^m, 2^{m+1})$ codes.
\end{itemize}

These codes are prototypes for higher rate dual-containing codes
described later.

These codes are given in terms of group ring codes over the group ring
of a direct product of cyclic groups. 
  
\subsubsection{Class 1}\label{sec:first}
The case $m=1$ is an $(8,4,4)$ self-dual code and is thus the Hamming $(8,4,4)$
code.
The other cases may be
considered as generalisations of  this to longer
lengths and increasing the distance. 
The case $m=2$ is an $(32,16,6)$ self-dual code.  
The best known binary $(32,16)$ codes 
 are $(32,16,8)$ codes and self-dual examples are known - see
 \cite{con1}, \cite{con2}.
The $(32,16,6)$ self-dual code fits into the general class and is
easy to describe and implement and is produced from an algebraic
formula. 
A   
binary $(32,16,8)$ self-dual code may however  be readily 
 constructed by the group ring
 methods but does not fit into this general class.

The next few cases in this class are  $(128, 64, 12), (512, 256,18), (2048,
 1024, 36)$  self-dual
codes. 
Examples beyond $(128,64,12)$ with described generator and
check matrices and known distances do not seem to be known. Another
 advantage is that the codes can be
algebraically stored   and  the matrices 
constructed algebraically as required, thus saving storage and power.

\subsubsection{Class 2}
Class 2 are   $(12,6,4), (72,36, 8), (432, 216, 16), (2592,
1296,32)$ etc.\ self-dual codes. $(12,6,4)$ is best possible although
$(72,36,8)$ is not.\footnote{The best known is $(72, 36,
  12)$.} Examples of this type    
beyond $(72,36)$ with described generator and
check matrices and known distances do not seem to be known. Again the
codes can be stored by an algebraic formula and the  matrices can be
constructed algebraically as required. 
 
\subsubsection{Golay}
The most famous self-dual code is the Golay $(24,12,8)$ code. This has
also been constructed by the group ring methods, \cite{ian}, as a
self-dual code in the group ring of the dihedral group. 
\subsubsection{Comparison}
 The Reed-Muller $(512, 256, 32)$ self-dual code may be compared with
 the $(512,256, 18)$ and $(432,216,16)$ codes here.
Although this  Reed-Muller code has better distance, the codes here have some
  other advantages. They   fit into the  more general picture and their
  generator and check matrices are easy to describe and
  construct. Moreover as they are algebraically produced they can be
  stored by an algebraic formula and reproduced  as
  needed, thus requiring low storage
  and low power. In further work, these codes are used to
  construct classes of convolutional codes. 


\subsection{Dihedral Self-Dual Codes}\label{sec:dih}
 Series of binary self-dual codes are  
 also obtained  by considering dihedral group ring codes and
 generalised dihedral group ring codes. Further  examples  from the
 general dihedral cases have still to be exploited.  

These dihedral-type  module codes show particular promise. 

By specifying certain {\em difference sets},  
 self-dual $(8m-2, 4m-1)$ codes
 are obtained when 
 $(4m-1)$ is a power of a prime and  $m$ odd. See \cite{lint} for
 definition and properties of difference sets.


$(22,11,6)$, $(38,19,8)$ and   $(54,27, 10)$ self-dual codes are
 obtained in 
this way from the \\ \mbox{$(11,5,2), (19,9,4), (27,13,6)$} difference sets.  Now
$(22,11, 6)$ and $(38,19, 8)$ are best possible for self-dual binary codes.  

In general the distance $d$ of the
 $(8m-2, 4m-1)$ self-dual dihedral code is almost certainly 
 $m+3$ and has been shown for a large number of cases. 
This would then  give
 a series of good  self-dual codes $(8m-2, 4m-1, m+3)$ in which the
 distance over length approaches $\frac{1}{8}$.

The distance can be  increased by increasing  the
length and get for example 
$(2\times {11}^m, {11}^m,2\times 3^m)$ self-dual codes. Here
multiplying the length by $11$ multiplies the distance by $3$. 

\subsection{Dual-containing codes}\label{sec:dualcontain1}
A {\em dual-containing code} $\C$ is a code such that its dual, $\C'$,
satisfies $\C' \subset \C$. Examples of these are   
important  for quantum codes, see \cite{calderbank} and \cite{mackay}.

Using a similar technique in the group rings as
with the  self-dual
codes, dual-containing codes of the form $(4m,3m)$
with rate $\frac{3}{4}$ are obtained. 
Specifically  $(2\ti 8^m, \frac{3}{2}\ti 8^m,
2^m)$  binary
dual-containing codes are obtained, giving for example $(16,12,2), (128, 96,4),
(1024,768,8)$ codes which are dual-containing.


Continuing thus,  $(8m,7m)$ codes of rate $\frac{7}{8}$ 
in a series of $(2\ti 16^m, \frac{7}{4}\ti 16^m, 2^m)$ dual-containing 
codes are obtained. 

 See also \sref{gf4} for dual-containing codes
over $GF(4)$. 

Quantum codes constructed  by the method of 
\cite{calderbank} 
using the constructions here are stated in \sref{quantum1} and are
further expanded on in \sref{quantum}.

Indeed it is possible to get $(16m,15m)$ etc.\ dual-containing 
 codes in a similar manner but 
details are not included.
\subsubsection{Further}
 $(4\ti 8^m, 3\ti 8^m,2^{m+1})$ dual-containing binary codes of rate
 $\frac{3}{4}$ are also
 obtained.  This
gives \\ $(32, 24, 4), (256, 192, 8), (2048, 1536, 16)$ etc.\
 dual-containing 
codes. The $(32,24,4)$ is best possible.
\subsubsection{Dihedral} 
The dihedral codes as described in \sref{dih} may also be extended to give $\frac{3}{4}$ rate
and higher rate 
dual-containing codes; this is the subject of further work. 

\subsubsection{Dual of the dual-containing}\label{sec:dualofdual}
The dual of the codes $(2\ti 8^m, \frac{3}{2}\ti 8^m,
2^m)$ in \sref{dualcontain1} are $(2\ti 8^m, \frac{1}{2}\ti 8^m)$
codes. These have rate
$\frac{1}{4}$, not so good,  but  
have  `nice' check elements/matrices. They are $(2\ti 8^m,
\frac{1}{2}\ti 8^m, 2\ti 4^{\frac{m+1}{2}}\ti 3^{\frac{m-1}{2}})$ codes for
      $m$ odd and  $(2\ti 8^m,
\frac{1}{2}\ti 8^m, 2\ti 4^{\frac{m}{2}}\ti 3^{\frac{m}{2}})$ codes for $m$
      even. 
 Similar results hold for duals of the other dual-containing codes
 presented here.


\subsection{Extending by Intertwining}\label{sec:ldpc1}

\subsubsection{Extension of self-dual}
Class 1  may  be extended or 
`intertwined' in a certain way
 to  obtain \\ \mbox{$(2\ti (4n)^m,
  (4n)^m, 2\ti 3^{\frac{m}{2}})$}, $m$ even, or 
$(2 \ti (4n)^m
, (4n)^m, 4 \ti
3^{\frac{m-1}{2}})$, $m$ odd,   codes for any $n\geq 1$.   Class 2 may be
extended or `intertwined' to obtain $(2\ti (6n)^m, (6n)^m,
 2^{m+1})$ codes, for any $n\geq 1$. 
For large $n$, and 
    $m$ small compared to $n$, these codes may be
    considered as {\em LDPC codes}. Notice that the distance depends
  on $m$ so,  as expected in an LDPC self-dual code, the distance will
  be small compared to the length. 

So for example self-dual codes of the types 
\mbox{$(8n, 4n, 4), (32n^2, 16n^2, 6),
(128n^3, 64n^3, 12)$} and \\
\mbox{ $(12n,6n,4),
(72n^2,36n^2, 8),  (432n^3, 216n^3, 16)$} are available.

Probably only the first two of each of
these series could, for large $n$, be considered for practical
purposes as `Low density'
since 
from then on  the density, although small compared to $n$, will be greater
than $20$. 
{\em However the 4-cycles are
determinable and these are `far apart'} in the sense that the indices
that make up any 4-cycle occur in rows at least $n$ apart.


As would be expected for self-dual LDPC codes, it is necessary
to increase the density in order to increase the distance.

The $(8n,4n,4)$ self-dual codes may be considered as an extension
of the Hamming $(8,4,4)$ code to low density cases.
   
\subsubsection{Dihedral expansion} The dihedral self-dual codes  as
described in \sref{dih} may
 also be intertwined to obtain higher length codes giving for
example $(22n,11n,  6), (38n, 19n,8), (54n,27n,10), (86n, 43n, 14)$ self-dual 
codes for any $n\geq 1$.  For $n$ large these could be considered as
LDPC codes. The  short 4-cycles are determinable and the indices that
occur in any 4-cycle are in rows which are at least $n$ places apart.

\subsubsection{Expansion by  intertwining of dual-containing.} 
As with the self-dual codes, 
the dual-containing may be expanded by {\em intertwining} 
to obtain other higher length dual-containing codes. Thus
for example $(2\ti (8n)^m, \frac{3}{2}
\ti (8n)^m, 2^m)$ and $(4\ti (8n)^m, 3\ti (8n)^m,2^{m+1})$
dual-containing binary codes of rate
 $\frac{3}{4}$  for any  $n,m$ are obtained. 


This gives  $(16n, 12n, 2), (128n^2, 96n^2, 4)$, 
  $(32n, 24n, 4), (256n^2, 192n^2, 8)$ etc.\
 dual-containing 
codes.  For $n$ large these could be
considered as LDPC dual-containing codes.  
Beyond these, although the density may be
small compared to the length, the density is probably too large to be
of  practical use as a low density code.
However the 4-cycles are determinable in all cases and 
these dual-containing codes have the property that the indices which
occur in any 4-cycle are in rows at least $n$ apart. 

Dual-containing codes   
$(32n, 24n,2), (512n^2, 384n^2,4)$ etc.\ codes with  rate
$\frac{7}{8}$ are also obtainable.

The intertwining
nature of the constructions suggests they 
should perform  better than the distances would indicate.  
\subsubsection{Expansion of dual of dual-containing.} The codes of
\sref{dualofdual},
$(2\ti 8^m,
\frac{1}{2}\ti 8^m, 2\ti 4^{\frac{m+1}{2}}\ti 3^{\frac{m-1}{2}})$
      , 
      $m$ odd and  $(2\ti 8^m,
\frac{1}{2}\ti, 2\ti 4^{\frac{m}{2}}\ti 3^{\frac{m}{2}})$, $m$
      even, may be intertwined to give $(2\ti (8n)^m,
\frac{1}{2}\ti (8n)^m, 2\ti 4^{\frac{m+1}{2}}\ti 3^{\frac{m-1}{2}})$ codes for
      $m$ odd and  $(2\ti (8n)^m,
\frac{1}{2}\ti (8n)^m, 2\ti 4^{\frac{m}{2}}\ti 3^{\frac{m}{2}})$ codes for $m$
      even.   

This gives $(16n, 4n, 8), (128n^2,32n^2, 24)$ codes. The density of
the check matrices, which are extremely nice, are respectively $4$ and
$10$. Thus these can be considered as LDPC codes for $n$
large. Moreover the symbols in any  4-cycle  are in rows at least $n$
places apart. 

\subsection{Self-dual codes and dual-containing codes over
  $GF(4)$.}\label{sec:gf4}  

The construction of \cite{calderbank} may be used to define quantum
  codes from dual-containing codes over $GF(4)$. The classes of
  quantum codes obtained from the dual-containing codes constructed
  here are described in \sref{quantum1}. 

 \cite{brower}  contains tables  of some known
quantum codes with best known distances up to length $128$.

The binary self-dual and dual containing codes
  as described in \sref{construct} and \sref{dualcontain1} 
may   be considered as codes over
  $GF(4)$ with the same length and distances.

The primitive element is
  not involved in these constructions. Using the primitive element and
  the  symplectic inner product, further classes  of self-dual
  and dual-containing codes over $GF(4)$ are obtained as follows:

\begin{enumerate}
\item $(2\times 4^m, 4^m, 2^{m+1})$ self-dual codes.
Thus we get $(8, 4, 4), (32, 16, 8), (128, 64, 16)$ etc. self-dual
codes over $GF(4)$.  
\item $(2\ti 4^m, \frac{3}{2}\ti 4^m, 2^{m})$ dual-containing  codes
  of rate $\frac{3}{4}$.  These are   $(8,6,2), (32,24,
  4), (128, 96,8)$ etc.\ dual-containing codes over $GF(4)$.
 
\item $(2\ti 8^m, \frac{7}{4}\ti 8^m, 2^{m})$ dual-containing  codes of
  rate $\frac{7}{8}$. These are  $(16, 14, 2), (128, 112, 4)$ etc.\
  dual-containing codes.
\item $(2\times (4n)^m, (4n)^m, 2^{m+1})$ self-dual codes.
These are  $(8n, 4n, 4), (32n^2, 16n^2, 8), (128n^3, 64n^3, 16)$
etc.\ self-dual codes over $GF(4)$ for any $n\geq 1$.  

\item $(2\ti (4n)^m, \frac{3}{2}\ti (4n)^m, 2^{m})$ dual-containing  codes
  of rate $\frac{3}{4}$. These are $(8n,6n,2), (32n^2,24n^2,4)$ 
etc.\ dual-containing codes for any $n\geq 1$.
 
\item $(2\ti (8n)^m, \frac{7}{4}\ti (8n)^m, 2^{m})$ dual-containing  code of
  rate $\frac{7}{8}$. These are  $(16n, 14n, 2), (128n^2, 112n^2, 4)$ etc.\
  dual-containing codes for any $n\geq 1$.
\item Higher rates may also be obtained.
\end{enumerate}

The  generator and check
matrices of these codes are easy to produce and can
be stored algebraically by formulae.

\subsection{Quantum codes}\label{sec:quantum1} By the construction of \cite{calderbank}
  the dual-containing codes obtained may be used to construct the following
  classes of quantum codes.

The codes over $GF(4)$ in \sref{gf4} may be used to construct the
following classes of quantum codes. 
\begin{enumerate}
\item The  $(2\times 4^m, 4^m, 2^{m+1})$ self-dual codes 
determine $[[2\ti 4^m, 0, 2^{m+1}]]$ quantum
codes, giving \\ $\mbox{[[8,0,4]], [[32,0,8]]}$ etc.\ quantum codes. Now
$[[8,0,4]]$ is best possible -- see \cite{brower}.  
\item $(2\ti 4^m, \frac{3}{2}\ti 4^m, 2^{m})$ dual-containing  codes
  of rate $\frac{3}{4}$ give rise to $[[2\ti 4^m,
 \frac{1}{2}\ti 4^m, 2^m]]$ quantum codes. Thus  $(8,6,2), (32,24,
  4), (128, 96,8)$ etc. dual-containing codes 
over $GF(4)$ produce  $[[8,4,2]], [[32,16, 4]],[[128,64,8]]$
  etc.\ quantum codes. Now $[[8,4,2]]$ is best possible - see \cite{brower}.
 
\item $(2\ti 8^m, \frac{7}{4}\ti 8^m, 2^{m})$ dual-containing  code of
  rate $\frac{7}{8}$ give rise to 
$[[2\ti 8^m, \frac{3}{2}\ti 8^m, 2^m]]$ quantum
codes. Thus we get $[[16,12,2]], [[128, 96, 4]]$ etc.\ quantum codes.
\item $(2\ti (4n)^m, \frac{3}{2}\ti (4n)^m, 2^{m})$ dual-containing  codes
  of rate $\frac{3}{4}$ give rise to $[[4n^m,
 \frac{1}{2}\ti 4n^m, 2^m]]$ quantum codes for any $n\geq 1$. 
Thus $(8n,6n,2), (32n^2,24n^2,4), (128n, 96n,8)$ etc. 
dual-containing codes 
over $GF(4)$ produce  $[[8n,4n,2]], [[32n^2,16n^2, 4]],[[128n^3,64n^3,8]]$
  etc.\ quantum codes of rate $\frac{1}{2}$. 
 
\item $(2\ti (8n)^m, \frac{7}{4}\ti (8n)^m, 2^{m})$ dual-containing  code of
  rate $\frac{7}{8}$ give rise to 
$[[2\ti (8n)^m, \frac{3}{2}\ti (8n)^m, 2^m]]$ quantum
codes for any $n\geq 1$. Thus we get $[[16n,12n,2]], [[128n^2, 96n^2, 4]]$
  etc.\ quantum codes of rate $\frac{3}{4}$. 
\item Higher rates may also be obtained.
\end{enumerate}
 
The binary dual-containing codes obtained in \sref{products},
\sref{dualcontain} and \sref{dualcontain8}
 give the following classes of quantum codes.    
\begin{itemize} 
\item The dual-containing binary codes $(2\ti 8^m, \frac{3}{2}\ti 8^m,
  2^m)$ of rate
$\frac{3}{4}$ may be used to obtain $[[2\ti 8^m,
    \frac{1}{2}\ti 8^m, 2^m]]$ quantum codes with rate $\frac{1}{2}$ 
by the construction of \cite{calderbank}.  This gives $[[16,8,2]],  
[[128,64,4]]$ etc.\ quantum codes of rate $\frac{1}{2}$.
\item The binary codes $(2\ti (16)^m,\frac{7}{4}\ti (16)^m, 2^m)$ of
rate $\frac{7}{8}$ give rise to $[[2\ti (16)^m,
    \frac{3}{2}\ti (16)^m, 2^m]]$ quantum codes of rate
$\frac{3}{4}$ by the construction of \cite{calderbank}.
 This gives $[[32, 24,2]], [[512, 384, 4]]$ etc.\ quantum
codes. 
\item $(2\ti (8n)^m, \frac{3}{2}\ti (8n)^m, 2^m)$ binary codes lead to
$[[2\ti 8n^m, \ti 8^nm, 2^m]]$ quantum codes for any  $n \geq 1$. 
This gives $[[16n,8n,2]],
[[128n,64n,4]]$ etc.\ quantum codes. 
\item $(2\ti (16)^mn,\frac{7}{4}\ti
(16)^mn, 2^m)$ binary  codes give   $[[2\ti 
(16n)^m, \frac{3}{2}\ti (16n)^m, 2^m]]$
quantum codes for any $n\geq 1$.
\item The self-dual codes $(2\ti 4^m,4^m,d)$ also determine $[[2\ti 4^m,0,d]]$
  quantum codes.
\item Higher rate binary codes dual-containing codes 
may also be constructed to produce  higher rate quantum codes. 
\end{itemize}

\subsection{Numbers} 
Consider $S = \{3^i + 3^{m-i}\}$ for $i = 0, 1, \ldots, m$. Then
the minimum of $S $ is $3^{\frac{m}{2}} + 3^{m-\frac{m}{2}} = 2 \ti
3^{\frac{m}{2}}$ when $m$ is even and is $3^{\frac{m-1}{2}} + 3^{m -
    \frac{m-1}{2}} = 3^{\frac{m-1}{2}} + 3^{\frac{m+1}{2}}=
3^{\frac{m-1}{2}}(1 + 3) = 4\ti 3^{\frac{m-1}{2}}$ when $m$ is
odd. These are the numbers which occur as distances in the first
class. The numbers that come in the second class are derived from
$ 2^i + 2^i = 2^{i+1} $.  
\subsection{Length and complexity} 
The codes here are produced algebraically and can be stored by an
algebraic formula. The matrices resulting from the algebraic
formulation in the group ring are easily programmed and reproduced as
needed and thus the codes  require low storage
and low power. 

The distances are also proved algebraically and a 
 required distance is obtainable at a desired rate by making the
 length large enough.
In general the calculation of distance is an NP-complete problem of
$O(2^r)$ where $r$ is the dimension of the code.

In cases where the groups are the direct product of cyclic groups, 
modified Discrete Fourier Transforms could be used to
  speed up the calculations if required. 
\subsection{Proofs} The proofs of the distances in general rely on finding a
certain type of 
distribution in the small cases within the group ring 
and then using properties of direct products 
to extend this to finding the smallest possible lengths/supports of the group
ring elements in the codes.


   

\subsection{Prime field dual-containing codes} The general techniques
described here can be applied to obtain self-dual and dual-containing
 codes over (other) prime fields. This is the subject of further work. 


\subsection{Isodual} Many of the constructions require an element to
 be symmetric. This condition can be relaxed and then {\em isodual codes} are
 obtained in place of self-dual codes and {\em codes containing a code
 equivalent to its dual} are obtained in place of dual-containing
 codes. Perhaps these latter codes could be called {\em
 isodual-containing codes}.  Relaxing the symmetric condition 
gives more codes and some with better distances; this is the
 subject of further investigation.

\section{Self-dual codes}\label{sec:general}

\subsection{Further notation and background}
$RG$ denotes the group ring of the  group $G$ over the ring
$R$. Further details on group rings may be obtained in \cite{seh}. Group ring
zero-divisor and group ring unit-derived codes are defined in
\cite{hur1} and the reader is referred to this paper for 
further notation and background.

$R_{n\ti n}$ denotes the ring of $n\ti n$ matrices over $R$.
If $u \in RG$ then $U \in R_{n\ti n}$ is
the image of $u$ under an isomorphism, $\phi$,  between $RG$ and the ring of
$RG$-matrices inside $R_{n\ti n}$ as given for example  in \cite{hur2}. 
The  $\rank u$ is defined to be $\rank U$.

The concept of `linear
independence' is often required and in these cases it is assumed that  $R$ is a
field. Many of the constructions however  can be formulated over
 systems other than fields.\footnote{The cases where $R$ is a group ring itself is closely
related to {\em convolutional codes}.}


\subsection{General formulation in group rings}\label{sec:self}
  
Form self-dual codes in $RG$ as follows. Suppose $|G| =m= 2q$ 
and that 
$u \in RG$ has the following properties:
\begin{enumerate}
\item $u^2 = 0$.
\item $u = u\T$.
\item  $\rank U = \rank u = q$.
\end{enumerate}

Then $u$ generates a self-dual code as follows. Consider $G = \{g_1,
g_2, \ldots , g_m\}$. Let $S= \{g_{i_1},
g_{i_2}, \ldots, g_{i_q}\}$ be chosen so  that $Su$ is linearly
independent. As pointed out in \cite{hur1} such a set always exists
since $\rank u  = q$. 
In most cases for  a natural ordering the set
 $S= \{ g_1,g_2, \ldots, g_q\}$, the first $q$ elements of $G$, is
such that $Su$ is linearly independent and in any
case by reordering the elements of $G$ it may be assumed that  $S$ consists of
the first $q$ elements. 
  
The self-dual code is then $\C = Wu$ where $W$ is the $R$-module generated
by $S$. A matrix version is obtained by applying the isomorphism
$\phi : RG \rightarrow R_{n\ti n}$ as explained in \cite{hur1}.

\subsubsection{Note 1}
Let $RG$ be a group ring and suppose that $\C$ is a code obtained 
with  $RG$-matrices $K,L$ satisfying  $KL = 0$ with $\rank K +
\rank L = n = |G|$.  The
code  is generated by $K$ and  `checked' by $L$. This is the case for 
example
with all cyclic codes which are zero-divisor group ring codes of the
cyclic group ring.   More precisely the matrix code is $\C = \al K$ where
$\al$ has length equal to $r=\rank K$, and the first $r$ rows of $K$
are linearly independent\footnote{A natural ordering
  of the elements of $G$ will almost always ensure this but in any case the
  elements of $G$ can be suitably ordered.}. Then $y\in C$ if and only if $yK = 0$ if and
only if  $L\T y\T = 0$. Thus $L\T$ is the check matrix in the
usual notation. 

The code is self-dual  if and only if $K = L\T$. If now $k,l\in RG$
are the elements corresponding to $K,L$ respectively we see that the
conditions for a self-dual code translates in the group ring setting
to finding an element $k\in RG$ such that $kk\T = 0$ and $\rank k =
\rank k\T = 1/2|G|$. Further if $k = k\T$ (i.e. $k$ and $K$ are 
 symmetric)  then the code is obtained from a group ring element $k$
 with   $k^2 =0$ and $\rank k= \rank K = 1/2|G|$. 

Thus in many situations, including (symmetric) cyclic codes,
self-dual codes are
obtained by the method of \sref{self}. It is more difficult to obtain
elements $k$ such that $kk\T =0$ with $\rank k = \rank k\T = 1/2|G| $
but  all  self-dual are obtainable this way when derived from group
rings; in particular cyclic self-dual codes and many other group ring
self-dual codes come about this way.  
\subsubsection{Note 2: Isodual codes} If the symmetric condition $u\T
= u$ is
omitted in \sref{self} then {\em isodual codes} are obtained. An isodual
code is a code equivalent to its dual. In this case we have $u^2 = 0,
\rank u = q = \rank u\T$. The check matrix is $U\T$ as opposed to $U$
in the self-dual case. However the group ring code determined by $u$
is equivalent to the group ring code determined by $u\T$; note 
from \cite{hur2} that
$u\T$ is the element obtained by interchanging the coefficients of
$g$ and $g^{-1}$ for every $g\in G$ in the expression for $u$ and that
if $U$ is the matrix of $u$ then $U\T$ is the matrix of $u\T$ in the
isomorphism between the group ring and the ring of matrices.

There is more freedom is the choice of $u$ if it is  not required that
$u$ be symmetric. Thus higher distance isodual codes may be obtained.

In other cases also it is possible to obtain codes which contain codes
equivalent to its dual. These should possibly be called
{\em isodual-containing codes}. 
\subsection{Matrices}

Consider now    $u\in RG $ which  has $RG$-matrix of the form 
$U = \left(\begin{array}{rr} A \\ B  \end{array}\right)$ where  
$\rank A  = \rank u = q$.
Then a generating matrix for the code is $A$. 
  
Suppose also $uv = 0$, $\rank v = m-q$ and $v$ has $RG$-matrix 
$\left( P , Q \right)$ where $\rank P = m-q$. Then a check matrix for
the code is $P\T$. 

If $U$ has the form   
$\left(\begin{array}{cc} I_q  & B \\ C & D \end{array}\right)$, the code
 then has generating matrix $\left( I_q, \, 
   B \right)$ which is in standard form.

Suppose then in the case of a self-dual code of \sref{self}  that $U =
\left(\begin{array}{ll} I_q & B \\ B & I_q \end{array}\right)$ with $u^2 =
0$ and $\rank u = q = \frac{m}{2}$. Then a generator matrix is $(I_q, B)$ and a
  check matrix is ${\left(\begin{array}{ll} I_q \\ B \end{array}\right)}^\T =
 (I_q , B\T)  $ and this is $(I_q, B)$ when $u$ is symmetric, as would
  be expected for a self-dual code.

The distance of the code is determined by $B$ essentially. 
\subsection{Listing of elements} A group ring code  is
independent  of the listing of the group elements. The corresponding
matrices depend on the listing but  equivalent matrix codes are
obtained by changing the listing. Adopt the following listing for a
direct product. Suppose a given  listing for $H$ is
$H= \{h_1, h_2, \ldots , h_r\}$ and a given listing for $K$ is $K = \{k_1,
k_2 \ldots, k_t\}$. The listing for $H\ti K$ is then taken to be 
$\{k_1H \cup k_2H \cup \ldots \cup k_tH\}$.
\subsection{Explicit groups}
Let  $G= H\times C_2$, where  $H= \{g_1, g_2, \ldots, g_q\}$ and 
$C_2 = \{1, h\}$.

Then $G$ is listed  by $G= \{g_1, g_2, \ldots , g_m, hg_1, hg_2, \ldots,
hg_q\}$.  The $RG$-matrix  is then 
$\left(\begin{array}{rr} A & B \\ B & A\end{array}\right)$ where
  $A,B$ are $RG$-matrices of $H$.
 
In the situation where $u = 1+
h(\di\sum_{j=1}^{m}\al_{j}g_{j})$ 
the matrix $U$ is  $\left(\begin{array}{ll} I_q & B \\ B &
  I_q\end{array}\right)$ and this has at least $\rank q$.  
If $u$ is symmetric then $U$ is a symmetric matrix (and so $B\T=B$)
and  a  self-dual code is obtained.

\subsection{Self-dual binary codes from direct products}\label{sec:products}
Let $G = C_4 \times C_4 \times \ldots \times C_4 \times C_2 =
C_4^m\times C_2$  where the $C_4$ are generated by $a_i$ for $i=1,\ldots,
m$ and $C_2$ is generated by $h$.

Then $|G| = 4^m\times 2$.

List the $i$th $C_4$ as $\{1, a_i,a_i^2,a_i^3\}$ and $C_2 = \{1,h\}$.

Then the listing  of $C_4^m$ is $L = \{1, a_1, a_1^2, a_1^3, a_2,
a_2a_1,a_2a_1^2,a_2a_1^3, a_2^2, a_2^2a_1, a_2^2a_1^2,a_2a_1^3,
\ldots, \ldots, a_m^3a_{m-1}^3\ldots a_1^3\}$ and the  listing of 
 $G$ is  $L\cup hL$.

Consider the group ring $\Z_2G$.

Define $u_1 = a_1+a_1^2+a_1^3$, $u_2 = u_1(a_2+a_2^2+a_2^3)$,
$\ldots$, $ u_m = u_{m-1}(a_m + a_m^2 + a_m^3)$.

Then it it easy to check that $u_i^2 = 1$. Also $u_i$ has $3^i$
distinct elements and is symmetric. 

Let $u= 1 +hu_m$. Then $u^2 = 1 +h^2u_m^2 = 1 + 1 = 0$.
Now $u$ has $1 + 3^m$ elements and is symmetric. Also $U$ has matrix
$\left(\begin{array}{rr} I &B \\ B & I\end{array}\right)$ 
for symmetric $4^m\times 4^m$
  matrix  $B$ with $B^2 =
  I$ and $I = I_{4^m}$. Thus $U$ and $u$ have $\rank 4^m$.

Since $u$ is symmetric and has rank $4^m$ it determines an $(4^m\times
2, 4^m)$ self-dual code $\C$. 

Let $S$ be the set of elements in $C_4^m$, the first $4^m$ elements of
$G$. The code is then generated by $Su$.

The generator matrix of the code is $(I,B)$ and the check matrix is
$\left(\begin{array}{r} I \\ B \end{array}\right)^{\T} = (I, B^{\T})$,
and as $u$ is symmetric,  $B = B^\T $.

It remains to determine the distance of the code.

\begin{theorem}\label{thm:distance} $\C$ has distance $2\times
  3^{\frac{m}{2}}$ when $m$ is even and has distance $4\times
  3^{\frac{m-1}{2}}$ when $m$ is odd.
\end{theorem}

The methods in the proof of \thmref{distance} are of general interest
and show how distances may be proved algebraically using group
rings. The {\em support of a group ring element} is the number of
non-zero coefficients in its expression as a group ring element.
 The distance of a  group ring code is the shortest support of any
group ring element in the code. 

Before proving the theorem  in general it is useful to look at
 some small cases.

The following Lemma is useful. 
\begin{lemma}{\label{lem:lem1}} 
Consider $u= 1 + h(a + a^2 + a^3)w$ and $T = \{u, ua,
  ua^2, ua^3\}$. Then a linear combination of one element in $T$ is
  $a^j + h(a^{j+1} + a^{j+2} + a^{j+3})w$, a linear combination of two
  elements from $T$ is  $a^i + a^j + h(a^p + a^q)w$
a linear combination of three elements of $T$ is  $a^i+a^j+a^k
  +h(a^p)w$ and  a linear combination of four elements of $T$ is 
$1 + a + a^2 + a^3 +
  h(1+a+a^2+a^3)w$ with 
 $0\leq i,j,k,p,q,\leq 3$,  $i,j,k$
  are distinct and $p,q$ are distinct. 
\end{lemma}
\begin{proof}
This can be proved directly by listing all the cases. Alternatively a
counting argument may be given.
\end{proof}

Consider $m=1$ in \thmref{distance}. Then $S= \{1,a_1,a_1^2,a_1^3\}$,
$u = 1+h(a_1+a_1^2+ a_1^3)$
and $Su$ consists of:
$$\begin{array}{ccc} u &=& 1 +h(a_1+a_1^2+a_1^3) \\ a_1u&=& a_1
  +h(a_1^2+a_1^3 + 1)
  \\ a_1^2u &= &a_1^2 + h(a_1^3+1 +a_1)\\ a_1^3u &= &a_1^3 + h(1+a_1
  +a_1^2) \end{array}$$

Note that each element of $Su$ contains a distinguishing element which 
does not occur in any other element of $Su$ and the other three
elements have two elements in common with the other elements of $Su$.
In any sum of elements of $Su$ the distinguishing elements survive. 
Thus if this sum contains more
than four  elements  at least 4 elements survive. If the sum contains
two elements then the two distinguishing elements survive and also two
more elements survive. If the sum contains three elements then the
three distinguishing elements survive plus one further element -- of
the nine other not necessarily distinct elements at least one is not
cancelled. Thus the distance of this code
is 4 and we have a $(8,4,4)$ code. 

This is the best distance for an $(8,4)$ (self-dual or otherwise) code. 

Consider now $m=2$. In this case we get a $(32,16)$ code.

We have $u_1 = a_1 + a_1^2+a_1^3$, $u_2 =
(a_2+a_2^2+a_2^3+a_2^3)u_1$, and $u= 1+hu_2$.  As already noted this 
determines a $(32,16)$ self-dual code. Note that $u$ has
$1+3^2$ elements.

Now 
$S= 1,a_1,a_1^2,a_1^3, a_2,a_2a_1, \ldots , a_2^3a_1^3$. The code is
generated by $Su$.

Separate
$S$ into 4 sets as follows: $S_1 = \{1, a_2, a_2^2,a_2^3\}$, $S_2 = \{a_1,
a_1a_2,a_1a_2^2, a_1a_2^3\}$, $S_3 = \{a_1^2,
a_1^2a_2,a_1^2a_2^2,a_1^2a_2^3\}$, $S_4 = \{a_1^3,a_1^3a_2,a_1^3a_2^2,
a_1^3a_2^3$\}.
 
Consider now a sum of elements in $Su$ which is then 
$ (\al_01 + \al_1 a_2 + \al_2a_2^2+ \al_3a_2^3 +\be_0a_1 +
\be_1a_1a_2 + \be_2a_1a_2^2 +\be_3a_1a_2^3 + \ga_0a_1^2,
+ \ga_1a_1^2a_2+\ga_2a_1^2a_2^2 + \ga_3a_1^2a_2^3 +
\de_0a_1^3+\de_1a_1^3a_2 +\de_2a_1^3a_2^2
+ \de_3a_1^3a_2^3)u$.

We use the notation $\fbox{i} + a$ in a group ring to denote the sum
of  $i$
independent non-zero terms added to an element $a$ which has terms
independent of the terms in $\fbox{i}$.

We can assume the coefficient, $\al_0$, of $1$ is non-zero.

Now  $(\al_01 + \al_1 a_2 + \al_2a_2^2+ \al_3a_2^3)u$ is $u$ if
only one coefficient (which is $\al_0$) is non-zero, is $\fbox{2} + h(u_1a_2^i + u_1a_2^j)$ for two
non-zero coefficients, is $\fbox{3} + h(u_1a_2^i)$ for 3 non-zero coefficients
and is $\fbox{4} + h(u_1(1 + a_2+a_2^2+a_2^3))$ for 4 non-zero coefficients.
The worst scenario is where we get $\fbox{3} + h(u_1a_2^i)$ which has 6
elements.

Similarly we get $0$ or $ua_2a_1^i$, $\fbox{2} + 
h(u_1a_2^i + u_1a_2^j)a_1^i$ for two
non-zero coefficients, is $\fbox{3}  + (u_1a_2^i)a_1^i$ for 3 non-zero 
coefficients
and is $\fbox{4}  + u_1(1 + a_2+a_2^2+a_2^3)a_1^i$ for 4 non-zero
coefficients.

All the elements here have none in common with the other type.

The worst scenario gives 6 elements and thus we have a $(32,16,6)$
code. 

The full weight distribution of the code may be obtained is a similar way.




 





Notation: Suppose $u = \di\sum_{i=1}^n \al_ig_i \in RG$. Then the
support of $u$, written $supp(u)$ is the number of non-zero $\al_i$.

To prove the general case we need the following.
 Consider $RH = R(G\times A)$ the group ring of the direct product of
 the groups $G$ and $A$. 

\begin{lemma}\label{lemref:support}
Suppose $u \in RG, w\in RA$. Then $supp(uw) =
supp(u)supp(w)$.
\end{lemma}  

The proof of this is straight forward and is omitted. 
\begin{lemma}\label{lemref:support2}
Let $a_1, a_2, \ldots , a_t$ be distinct elements of $A$ and 
$u= \al_1u_1a_1 + \al_2u_2a_2 +\ldots + \al_tu_ta_t$ where $\al_i \not
= 0$ and $u_i \in RG$. Then $supp(u) = supp(u_1) + supp(u_2) + \ldots
supp(u_t) $. 
\end{lemma} 

We now prove \thmref{distance} in general. Let the 
 element $u$  be as defined. The ideas in the proof below may be used in
 other cases to prove the minimum distance for group ring codes.

\begin{proof} Let $G_n$ denote the direct product of $n$ copies of
  $C_4$. We are considering $G= G_n\times C_2$ and the group ring
  $\Z_2G$. We already know that $u^2 =0$, that $\rank u =1/2|G|$ and
  that $u$ generates a self-dual code. Set $S$ to be the elements  of
  $G_n$ and $W$ the module generated by $S$. The code then is
  $Wu$. Thus we need to show that the smallest support/ 
length of an element in $Wu$ is as stated. 

Let $K$ denote the cyclic group of order $2$ generated by $h$.
We have already seen that $(C_4\times K)u$ has distribution $ x
+ hy$ where $|x| =1, |y| =3, (ii) |x| = 2, |y| = 2 (iii) |x| = 3, |y|
= 1 (iv) |x|= 4, |y| = 4.$

Say the element $x+hy \in G\ti C_2$ has weight distribution
$(|x|,|y|)$.

Now $G = G_{n-1}\times C_4 \times K$ where $C_4$ is $\{1,a_n, a_n^2,
a_n^3\} =\{ 1,a,a^2,a^3\}$ say. 

Now $RG = (RG_{n-1})\times C_4 \times K$. Every element in
$RG_{n-1}\times C_4$ can be written in the form $q(\al_0 + \al_11a+\al_2a^2 +
\al_3a^3)$ with $q \in RG_{n-1}$ and the $\al_i = 0,1$. For a non-zero
element at least one of the $\al_i \not = 0$. Thus every non-zero element in
the code can be written in the form 

$w = q(\al_0 + \al_11a+\al_2a^2 + \al_3a^3)(1 + hu_n)$. 

Case 1. One of the $\al_1 \not = 0$ (and others all zero). Then $w =
qa^i + hqu_{n-1}(a^{1+i} + a^{2+i} + a^{3+i})$. This has $|q|  +
3|qu_{n-1}| $ elements.

Case 2: Two of the $\al_i \not = 0$. Then $w = 
q(a^i + a^j) + h(qu_{n-1}(a^k + a^l)$. This has $2|q| + 2|qu_{n-1}|$
elements.

Case 3: Three of the $\al_i \not = 0$. Then $w = q(a^i+a^j+a^k) +
h(qu_{n-1}a^t)$. (Where of course $i,j,k$ are all different.) This has
$3|q| + |qu_{n-1}|$ elements. 

Case 4: All of the $\al_i$ are non-zero. Then $w = q(1 + a+a^2+a^3) +
h qu_{n-1}(1+a+a^2+a^3)$. This has $4|q| + 4|qu_{n-1}|$ elements.

We can now decide the minimum weight/support by considering the
minimum which can occur in these 4 cases. We know by induction the
minimum of $|q|$ and $|qu_{n-1}|$. 

 We have shown that the distribution is $(1,3), (2,2),
(3,1), (4,4)$ when $n=1$ with minimum $(1,3), (3,1)$. Thus the next
minimum distribution is $(3,3), (3,3)$ which gives distance $6$ as
required. This is $n=2$ even. 

The next minimum distribution is $(3,9), (9,3)$ giving a distance of
$12$ as required.  This is $n=3$ odd.

Suppose the minimum distribution for $k$ even is $(3^{\frac{k}{2}},
3^{\frac{k}{2}}), (3^{\frac{k}{2}},3^{\frac{k}{2}})$ and is 
$(3^{\frac{k+1}{2}}, 3^{\frac{k-1}{2}}), (3^{\frac{k-1}{2}},
  3^{\frac{k+1}{2}})$ for $k$ odd.  Then the next minimum distribution is  
$(3^{\frac{k+2}{2}}, 3^{\frac{k}{2}}),
      (3^{\frac{k}{2}},3^{\frac{k+2}{2}})$  for $k$
	  even (or $k+1$ odd) and is $(3^{\frac{k+1}{2}},
  3^{\frac{k+1}{2}}), (3^{\frac{k+1}{2}},3^{\frac{k-1}{2}})$ for $k$
  odd (or $k+1$ even).

It is clear that for other non-minimal distributions at the $k$th
stage will give higher distributions at the $k+1$st stage.

Thus in general the minimum distribution is $(3^{\frac{n}{2}},
3^{\frac{n}{2}}), (3^{\frac{n}{2}},3^{\frac{n}{2}})$ for $n$ even
  giving a distance of $2\times 3^{\frac{n-1}{2}}$ and is 
$(3^{\frac{n+1}{2}}, 3^{\frac{n-1}{2}}), (3^{\frac{n-1}{2}},
  3^{\frac{n+1}{2}})$ for $n$  odd 
giving a distance of $3^{\frac{n-1}{2}}
    +  3^{\frac{n+1}{2}}
      = 3^{\frac{n-1}{2}}(1 + 3) = 4\times 3^{\frac{n-1}{2}} $.

\end{proof}
\subsection{Further self-dual codes from direct products}

Now consider $\Z_2G$ where $G = C_6^m \times C_2$. Suppose the cyclic
groups of order $6$ are generated by $a_i$ for $ i =1 , \ldots , m$
and $C_2 $ is generated by $h$.

Define $u_1= a_1+ a_1^2 + a_1^3 + a_1^4+ a_1^5$ ;  $u_2= (a_2+ a_2^2 +
a_2^3  + a_2^4 + a_2^5)u_1$ ;  $u_m= (a_m+ a_m^2 + a_m^3 + a_m^4+
a_m^5)u_{m-1}$.

Then $u_i^2 = 1$ for each $i$ and $u_i$ is symmetric with $5^i$
elements. Set $u = 1 + hu_m$. Then $u^2 = 0$, is symmetric  and $u$ has matrix 
$U = \left(\begin{array}{rr} I &B \\ B & I\end{array}\right)$ 
for symmetric $6^m\times 6^m$
  matrix  $B$ with $B^2 =
  1$ and $I = I_{6^m}$. Thus $U$ and $u$ have $\rank 6^m$.

Hence $u$ defines a $(2\times 6^m, 6^m)$ self-dual code $\C_m$.

It remains to determine the distance of $\C$.
\begin{theorem}\label{thmref:6power}$\C_m$ has distance $2^{m+1}$.
\end{theorem}
\begin{proof} The proof of this is very similar to the proof of
  \thmref{distance}. It depends on the following first case
  situation. Let $C_6$ be
generated by $a$ and define $ u_1 = a + a^2+a^3+a^4 +a^5$.
Then consider combinations over  $\Z_2$ of $u_1, u_1a, u_1a^2,
u_1a^3,u_1a^4,u_1a^5$.   

For one non-zero coefficient in the combination we get $u_ia^i$ which
has $5$ non-zero $a_i$, for two non-zero coefficients we get $a^i +
a^j$, for three  we get $a^i+a^j+a^k$, for four we get
$a^i+a^j+a^k+a^l$, for five we get $a^i$, and  for six  we get
$a^i+a^j+a^k+a^l+a^m +a^p$. 
The worst case in a sense is where we take five non-zero coefficients 
 and get just one $a^i$. However it does take five non-zero elements to
get this low distance. In all other cases we get at least two $a^i$. 
 This distribution is then extended to the general case as in 
\thmref{distance}.
\end{proof}

Thus we get $(12, 6, 4), (72,36, 8), (432,216, 16)$ etc.\ codes. Now
$(12,6,4)$ is best possible distance for a $(12,6)$ code. A $(72,36)$
self-dual binary code has best possible known distance of  $12$. 
 Now $(72,36,8)$ has certain advantages; as well as fitting into the
general infinite class and leading to extensions and intertwining with
 higher lengths,  its method of construction  enables a full
 description of its weight distribution.

\section{Self-dual codes from dihedral groups}\label{sec:dihedral} 

Let $D_{2m} = \, <a,b| a^m=1=b^2, a^b=a^{-1}>$ be the dihedral group of
order $2m$. A listing of the group is $D_{2m} = \{1, a, \ldots,
a^{m-1}, b, ba, \ldots, ba^{m-1}\}$. Any element in the group ring
$RD_{2m}$ can be given as $u =\di\sum_{i=0}^{m-1}\al_ia^i +
b\di\sum_{j=1}^{m-1} \be_j a^j$. 

Now consider $u= 1 + b\di\sum_{j=1}^{r}a^{t_j}= 1 + bD$
say. Then $u$ is automatically symmetric as $(ba^j)^{-1} = ba^j$. 
 Now work in characteristic 2. Then $u^2 = 1 + bDbD = 1 +
b^2D^{-1}D = 1 + D^{-1}D$. Consider situations when  $D^{-1}D =1$ and 
 then $u^2= 1+1 = 0$.   

In  this situation, $u$ generates a self-dual $(2m, m)$ code. 
Its matrix is of the form $(I_m, C)$ where $C$ is the reverse circulant
matrix corresponding to $D$. 

By abuse of notation let $D$ denote the group ring element and 
the set of group elements which occur in $D$.
Now consider suitable sets   $D$ for which $D^{-1}D = 1$.

These sets remind us of {\em difference sets} in groups, see
\cite{lint}. If $(v,k,\lambda)$ is a difference set in a group and $D$
is the corresponding group ring element in the group $G$ 
then (actually iff) $D^{-1}D = n + \lambda G$ where $n= k- \lambda$. Thus if
$\lambda$ is even and $n$ is odd then in characteristic 2, $D^{-1}D =
1$.  

In particular there exist $(4n-1, 2n-1, n-1)$ difference sets in the
multiplicative (cyclic) group of the field $F_{4n-1}$ when $4n-1$ is
a power of a prime. Thus when $n$ is odd  we
get from these self-dual $(2m,m)$ dihedral codes, where $m=4n-1$. This
gives for example $(22,11), (38,19), (54, 27)$ etc.\ self-dual codes. 

The distances of these codes are quite good and  we get 
$(22,11, 6), (38,19,8)$ self-dual codes.   

 In general the distance of  the  $(8n-2, 4n-1)$ code is probably 
$n+3$; this would give  a series of  $(2p,p,d)$ 
self-dual codes 
where $\lim_{p\rightarrow  \infty}\frac{d}{2p}  = \frac{1}{8}$ giving 
a series of `good' codes. 

In general {\em  sets of differences} in a group with particular properties are
needed and not necessarily difference sets. In order to define a
self-dual code it is sufficient that 
the set of differences contains each
difference an even number of times and that the difference set itself
has an odd number of elements; this ensures that $D^{-1}D =1$ in
characteristic 2.

The method  can also be applied to generalised dihedral groups. Let $G$ be
any abelian group. Then the generalised dihedral group,  written $Dih(G)$, 
is the semidirect product of $G$ and $C_2$, with $C_2$ acting on $G$
 by inverting elements. 

Every element in the group ring $RDih(G)$ may be written $u = P + bD$
 with $P,D 
\in RG, b^2 = 1$ and $ D^b = D^{-1}$. If $u= 1 +bD$ then in
characteristic 2, 
$u^2 = 1 + bDbD = 1 + D^{-1}D$.

Let $D$ be a difference set in $C_m$ and  consider $C_m^t$. 
In $Dih(C_m^t)$ define $u = 1 + b(D_1D_2\ldots D_t)$ with $b^2=1$ and
$b$ acting by inverting elements, where $D_i$ corresponds to $D$ in
the $i^{th}$ term of the direct product. Then in characteristic 2, $u = 1 +
bD_1D_2\ldots D_tbD_1D_2\ldots D_t = 1 + D_1^{-1}D_1D_2^{-1}D_2\ldots
D_t^{-1}D_t = 1 + 1 = 0$. Thus $u$ generates a self-dual $(2\times
m^t, m^t)$ code. 


This could be further generalised by considering suitable difference
sets $D_i$ in cyclic groups $C_{t_i}$ for $ i= 1,.., t$, forming $G = \prod C_{t_1}$ and
looking at $u = 1 + b(D_1D_2\ldots D_t$). 

There are other  possibilities in the dihedral and generalised dihedral
still to be exploited and studied.

\section{Expansion by intertwining} 
Let $G = 
C_{4n}^m\times C_2$  where the $C_{4n}$ are generated by $a_i$ for $i=1,\ldots,
m$ and $C_2$ is generated by $h$.
Then $|G| = (4n)^m\times 2$.

List the elements of $C_{4n}^m$ in the natural way as $L = 
\{1, a_1, a_1^2, \ldots, a_1^{4n-1}, a_2,
a_2a_1,a_2a_1^2,\ldots , a_2a_1^{4n-1}, \\ a_2^2, a_2^2a_1, a_2^2a_1^2,\ldots
a_2^2a_1^{4n-1},
\ldots,\ldots,  a_m^{4n-1}a_{m-1}^{4n-1}\ldots a_1^{4n-1}\}$ 
and then list $G$ as $L\cup hL$.

Consider the group ring $\Z_2G$.

Define $u_1 = a_1^n+a_1^{2n}+a_1^{3n}$, $u_2 = u_1(a_2^n+a_2^{2n}+a_2^{3n})$,
$\ldots$, $ u_m = u_{m-1}(a_m^n + a_m^{2n} + a_m^{3n})$.

Then it it easy to check that $u_i^2 = 1$. Also $u_i$ has $3^i$
distinct elements and is symmetric. 

Let $u= 1 +hu_m$. Then $u^2 = 1 +h^2u_m^2 = 1 + 1 = 0$.
Now $u$ has $1 + 3^m$ elements and is symmetric. Also $U$ has matrix
$\left(\begin{array}{rr} I &B \\ B & I\end{array}\right)$ 
for symmetric $4^m\times 4^m$
  matrix  $B$ with $B^2 =
  1$ and $I = I_{(4n)^m}$. Thus $U$ and $u$ have $\rank (4n)^m$.

Since $u$ is symmetric and has rank $(4n)^m$ it determines an $((4n)^m\times
2, 4n^m)$ self-dual code $\C$.

Let $S$ be the set of elements in $C_{4n}^m$, the first $(4n)^m$ elements of
$G$. The code is then generated by $Su$.

The generator matrix of the code is $(I,B)$ and the check matrix is
$\left(\begin{array}{r} I \\ B \end{array}\right)^{\T} = (I, B^{\T})$,
and as $u$ is symmetric,  $B = B^\T $.

The group ring elements are intertwined and the elements $ua^j$ and
$ua^k$ have elements in common if and only if $j \equiv k \mod
n$. Thus the following theorem follows directly as
in \thmref{distance}. 

\begin{theorem} The codes are $((4n)^m\times
2, (4n)^m, 2\ti 3^{\frac{m}{2}})$
 self-dual code for $m$ even and are $((4n)^m\times
2, (4n)^m, 4\ti 3^{\frac{m-1}{2}})$ self-dual code for $m$ odd.
\end{theorem} 

The `LDPC' codes do contain (short) 4-cycles but these cycles occur far
apart --  the indices involved in any 4-cycle are in rows of 
 length $n$ from one another. 
  
\section{Dual-containing codes of rate $\frac{3}{4}$}\label{sec:dualcontain} 

\subsection{General set-up} Consider $RG$ with $|G| = m = 4q$ 
which has an element $u$ such that :
\begin{enumerate}
\item $u^4 = 0$.
\item $u$ is symmetric.
\item $u$ and $U$ have $\rank  = 3q$.

\end{enumerate}
Since $u$ has $\rank 3q$ it  will follow that $u^3$ has $\rank q$. 
To show this we need the following well-known result on ranks of
matrices.
\begin{lemma}\label{lem:rank}
Suppose $A,B$ are $n\times n$ matrices. Then

(i) $\rank AB \leq \text{min}\{\rank A, \rank B\}$.

(ii) $\rank AB \geq \rank A + \rank B - n$.
\end{lemma} 

Then:
\begin{lemma}\label{lem:rank2} Suppose $U$ is an $4q\times 4q$ matrix
  with  $U^4 = 0$ and
  such that $U$ has $\rank 3q$. Then $\rank U^3 = q$.
\end{lemma}
\begin{proof}
Since $UU^3 =0$, $U^3$ is in the null-space of $U$ and so cannot have
rank greater than $q$. 
Now by \lemref{rank}, $\rank U^2 \geq 3q + 3q - 4q = 2q$.
Then again by \lemref{rank}, $\rank U^3 = \rank U^2U \geq 2q + 3q - 4q
= q$. Hence $\rank U^3 = q$.

\end{proof}

Thus $u$ will generate a dual-containing code $(4q,3q)$ of rate
$\frac{3}{4}$. $u$ is the generating element of the code and $u^3$ is
the check element. The generator matrix of the equivalent matrix code
is $U$ and only the first three-quarters of the rows of $U$
need be used. $U^3$ is the check matrix and only the first quarter of
its rows need be used as the check matrix.

The code is  $Wu$ where $W$ is a certain submodule which can be taken
to be the submodule generated by the first $3q$ elements of $G$. The
dual code is $Wu^3$ which  is obviously
contained in $Wu$; note that $Wu = RGu$ and also $Wu^3 =RGu^3$.

As is shown in \cite{hur1}, it is possible to obtain a submodule $W$ 
generated by
$S =\{g_{i_1}, g_{i_2}, \ldots , g_{i_{3q}}\}$ such that $Su$ is linearly
  independent. Then the code is $Wu$. By reordering if necessary it is
  possible to choose $S = \{1= g_1, g_2, \ldots , g_{3q}\}$.  Provided
  the first $3q$ rows of
  $U$ are linearly independent we may choose $S$ to be the first $3q$
  elements of $G$.

Relaxing the symmetric condition gives codes which
contain a code equivalent to its dual, {\em isodual-containing
  codes}. This gives  many more examples. In these situations we have
 $u^4 = 0$, $\rank u = 3q$ and $(u^3)^T$ is
the check element. Then the code contains a code which is equivalent
to its dual. 
  
\subsection{Explicit groups}

Suppose now a group ring $RG$  has an element $w$ with $w^4  =
1$. 
 Then  consider
$R(G\times C_2)$ where $C_2$ is generated by $h$. List the group
 elements by
$\{1, g_1, g_2, \ldots , g_n, h, hg_1, \ldots , hg_m\}$  and let $u = 1
 +hw$. 

 Then $u^4 = 0$ when  $R$ has characteristic $2$.  If $\rank
 u = \frac{3}{4}m$ then $\rank u^3 = \frac{1}{4}m$ so that $u^3$ is
 then the check  element. The code is given by $\C = \{\be u\}$.
 
If also $u$ is symmetric then we get a dual-containing code as the
dual code is the set of all $\be u^3$ which is  $ \be u^2\times u \in \C$. 

Using this we  now produce such codes explicitly and derive  
their distances.

 Let $ H = C_8^m$ and consider the group $C_8^m\times
 C_2$ which has order $2 \times 8^m$.

Consider $u_1 = a_1 + a_1^4 + a_1^7$. This is symmetric and satisfies
$u_1^4 = 1$. Let $u_2 = (a_2 + a_2^4 +a_2^7)u_1$ and in general $u_i =
(a_i+a_i^4 + a_i^7)u_{i-1}$. Then $u_i^4 =1$, $u_i$ is symmetric and
has $3^i$ elements. In particular $u_m$ has order $4$, thus 
invertible,  and is symmetric.

Define $u = 1 + hu_m$. Then $u^4 = 0$. We need to determine the rank
of $u$. Now the matrix of $u$ is $\left(\begin{array}{rr} I & B \\ B &
    I \end{array}\right)$ which by row operations is equivalent to $
\left(\begin{array}{rr} I & B \\ 0 & I + B^2\end{array}\right)$. Here
  $I$ and $B$ have size $8^m$ and $\rank I = 8^m$.

 To show that $u$ has
  $\rank =\frac{3}{4}\times (8^m\times 2)$ we now need to show that $I+B^2$ has
  $\rank =\frac{1}{4} \times (8^m\times 2)= \frac{1}{2}\ti 8^m$.   

It is clear that $B$ corresponds to the group ring element $u_m$ with
$B^4 = 1$ and $(I+B^2)^2 = 0$. Also $1+B^2$ correspond to $1+u_m^2$ which
is $1+(1+a_1^2 + a_1^6)(1+a_2^2+a_2^6)\ldots (1+a_m^2+a_m^6)$. 
Thus $I+B^2$ has the form $\left( \begin{array}{rr} P & Q \\ Q & P
\end{array}\right)$ where $P$ is non-singular and so $I+B^2$ has 
rank $\frac{1}{2}\ti 8^m$ as required.

Consider now $R=\Z_2$.

\begin{theorem} The code $\C$ has distance $2^m$.
\end{theorem}
\begin{proof}
The proof is very similar to the proofs for the self-dual
codes. The case $m=1$ has distance $2$ and this is extended to higher
$m$ using properties of direct products.
\end{proof}

Thus $(2\times 8^m, \frac{3}{2}\ti 8^m, 2^m)$ dual-containing codes 
are obtained.  For $m=1$ this is
an $(16,12,2)$ which is best possible. 
Next is  $(128,96,4)$, then  $(1024,768,8)$, $(8192,7168, 16)$ etc. 

The generator and check matrices for these codes are
immediately obtained from the group ring elements.

\subsubsection{Further} Consider the group ring $\Z_2G$ with $G = C_4
\cross C_{8}^m$, where $C_4$ is generated by $h$ and the $C_8$ are
generated by $a_i, i=1,2, \ldots, m$. As before define  $u_1 = a_1 + a_1^4 +
a_1^7$ and in general $u_i =
(a_i+a_i^4 + a_i^7)u_{i-1}$. Then $u_i^4 =1$, $u_i$ is symmetric and
has $3^i$ elements. In particular $u_m$ has order $4$ and is symmetric.

Define $u = 1 + h^2u_m$. Then $u$ is symmetric and 
 $u^4 = 1 + u_n^4 = 1 + 1 = 0$. The matrix
$U$ of $u$ has the form $\left(\begin{array}{rrrr} I & 0 & B&0 \\ 0
  &I&0&B \\ B&0&I&0 \\ 0 &B&0&I \end{array}\right)$. 

This by (block) row operations is row equivalent to

$\left(\begin{array}{cccc} 
I & 0 & B&0 \\ 0
  &I&0&B \\ 0&0&I+B^2&0 \\ 0&0&0&I+B^2 \end{array}\right) $. 

Now each
$I+B^2$ has rank $\frac{1}{2}\times 8^m$ and so $\rank U
= 8^m+8^m+\frac{1}{2}\times + 8^m\frac{1}{2}\times 8^m = 3\times 8^m$.  


It
follows that $\rank U^3 = 8^m$. Thus we get a $(4\ti 8^m, 3\ti 8^m)$
code with generator element $u$ in the group ring or $U$ in the matrix
code and check element $u^3$ in the group ring code and $U^3$ in the matrix
code.

Call this code $\C_m$. 

\begin{theorem}$\C_m$ has distance $2^{m+1}$.
\end{theorem}

We thus get $(4\ti 8^m, 3\ti 8^m,2^{m+1})$ dual containing codes.

 This
gives $(32, 24, 4), (256, 192, 8), (2048, 1536, 16)$ etc.\
codes. The $(32,24,4)$ is best possible.
\section{Intertwining: Lengthening the dual-containing of rate
  $\frac{3}{4}$}

The above \sref{dualcontain} can be extended to give
longer  dual-containing codes in an intertwining way.

 Let $ H = C_{8n}^m$ and so we are considering the group $C_{8n}^m\times
 C_2$ which has order $2 \times (8n)^m$.

Consider $u_1 = a_1^n + a_1^{4n} + a_1^{7n}$. This is symmetric and satisfies
$u_1^4 = 1$. Let $u_2 = (a_2^n + a_2^{4n} +a_2^{7n})u_1$ and in general $u_i =
(a_i^n+a_i^{4n} + a_i^{7n})u_{i-1}$. Then $u_i^4 =1$, $u_i$ is symmetric and
has $3^i$ elements. In particular $u_m$ has order $4$ and is symmetric.

Define $u = 1 + hu_m$. Then $u^4 = 0$. We need to determine the rank
of $u$. Now the matrix of $u$ is $\left(\begin{array}{rr} I & B \\ B &
    I \end{array}\right)$ which by row operations is equivalent to $
\left(\begin{array}{rr} I & B \\ 0 & I + B^2\end{array}\right)$. Here
  $I$ and $B$ have size $(8n)^m$ and $\rank I = (8n)^m$.

 To show that $u$ has
  $\rank =\frac{3}{4}\times ((8n)^m\times 2)$ it is necessary 
 to show that $\rank (I+B^2) =\frac{1}{4} \times ((8n)^m\times 2)$.
  This is done as previously by looking at the group ring element
  corresponding to $I+B^2$.


It is also noted as before that $ua^j$ and $ua^k$ have elements in
common if and only if $j\equiv k \mod n$ so that the code is
intertwined. The distance then is precisely the same as for $n=1$. 
\section{Further: Dual-containing codes of rate $\frac{7}{8}$}\label{sec:dualcontain8}
The above \sref{dualcontain} can be generalised further.
 
Suppose there exists an element $u\in RG$ where $|G| = m = 8q$ with 

(i) $u^8=0$ and

(ii)$\rank u = 7q$. 

It then follows as in \lemref{rank2} that $\rank u^7 =
q$. 

We thus get a $(8q,7q)$ code generated by $u$ with check element
$u^7$. If further $u$ is symmetric this will be a dual-containing
$(8q, 7q)$ code. The code is given by $Wu$ for a module $W$ of rank
$7m$ and the dual is $Wu^3$ which is clearly contained in $Wu$.    

We produce explicit examples as follows:

 Let $ H = C_{16}^m$ and consider  the group $C_{16}^m\times
 C_2$ which has order $2 \times 16^m$.

Let $R = \Z_2$.

Consider $u_1 = a_1 + a_1^8 + a_1^{15}$. This is symmetric and satisfies
$u_1^8 = 1$. Let $u_2 = (a_2 + a_2^8 +a_2^{15})u_1$ and in general $u_i =
(a_i+a_i^8 + a_i^{15})u_{i-1}$. Then $u_i^8 =1$, $u_i$ is symmetric and
has $3^i$ elements. In particular $u_m $ order $8$ and is symmetric.

Define $u = 1 + hu_m$. Then $u^8 = 0$. We need to determine the rank
of $u$. Now the matrix of $u$ is $\left(\begin{array}{rr} I & B \\ B &
    I \end{array}\right)$ which by (block) row operations is equivalent to $
\left(\begin{array}{rr} I & B \\ 0 & I + B^2\end{array}\right)$. Here
  $I$ and $B$ have size ${16}^m$. Clearly then $\rank I = {16}^m$. 

 To show that 
  $\rank u = \frac{7}{8} \times {16}^n \times 2 $ it is necessary  to show
  that 
  $\rank I+B^2 =\frac{3}{8} \times {16}^m\times 2$.   

The details on this are omitted and consists of looking at the group
ring element corresponding to $I + B^2$ and showing it has the
required rank. 

\begin{theorem} These codes are $(2\times 16^m, \frac{7}{4}\ti 16^m,
  2^m)$ dual containing
codes of rate $\frac{7}{8}$. 
\end{theorem}

The first two  are $(32,28, 2), (512,448,4)$ dual-containing codes.

\section{Generally} 

Higher rate dual-containing codes may be obtained as follows. Only a
bare outline is given and details are omitted.

Let $|G| = 2^tq $ and  suppose we have an element  $u\in RG$
such that: 

\begin{enumerate} \item $u^{2^t} = 0$
\item $u$ is symmetric,
\item $\rank u = (2^{t} - 1)q$. 
\end{enumerate}

It will follow as before that $\rank u^{2^t-1} = q$. Thus we get 
$uu^{2^t-1}= 0$
where $\rank u = (2^t-1)q$ and $\rank u^{2^t-1} = q$. 
Then $u$ will generate a dual-containing code of type $(2^tq,
(2^t-1)q)$ so that the rate of the code is $\frac{(2^t-1)q}{2^tq} =
\frac{2^t-1}{2^t}$. The check element is $u^{2^t-1}$.

To get explicit examples consider: 
 Let $ H = C_{2^{t+1}}^m$ and consider  the group $C_{2^{t+1}}^m\times
 C_2$ which has order $2 \times 2^{(t+1)m}$.

In the group ring $RG$ assume $R$ has characteristic $2$.

Consider $u_1 = a_1 + a_1^{2^t} + a_1^{2^{t+1}-1}$. 
This is symmetric and satisfies
$u_1^{2^t} = 1$. Let $u_2 = (a_2 + a_2^{2^t} +a_2^{2^{t+1}-1})u_1$ and in
general $u_i =
(a_i+a_i^{2t} + a_i^{2^{t+1}-1})u_{i-1}$. Then $u_i^{2^t} =1$, $u_i$
is symmetric and
has $3^i$ elements. In particular $u_m $ has  
order $2^t$ and is symmetric.

Define $u = 1 + hu_m$. Then $u^{2^t} = 0$. 
We need to determine the rank
of $u$. Now the matrix of $u$ is $\left(\begin{array}{rr} I & B \\ B &
    I \end{array}\right)$ which by row operations is equivalent to $
\left(\begin{array}{rr} I & B \\ 0 & I + B^2\end{array}\right)$. Here
  $I$ and $B$ have size ${2^t}$. 
 To show that $u$ has
  $\rank  = (2^{t-1})2$ it is necessary to show that $I+B^2$ has
  $\rank = 2(2^t-1)-2^t$.   

\subsection{Intertwining to obtain higher lengths}
It is clear also as previously that the codes of \sref{dualcontain8}
 may be intertwined to obtain
  $(2\times (16n)^m,  \frac{7}{4}\ti (16n)^m,
  2^m)$ dual containing
codes of rate $\frac{7}{8}$ for any $n\geq 1$. 

\section{Self-dual and dual-containing codes over $GF(4)$}\label{sec:quantum} 
In \sref{products}, \sref{dualcontain} and \sref{dualcontain8}
 self-dual and dual-containing binary codes are obtained by
 considering the direct product of groups. These can be considered as codes over $GF(4)$.
 They can be modified as follows to give codes over $GF(4)$ with
 better distance. Let $\om$
 be the primitive element in $GF(4)$.

Define $u_1 = \om a_1+a_2^2+\om a_1^3$ and in general $u_{i+1} =
u_i(\om a_{i+1} + a_{i+1}^2 + \om a_{i+1}^3)$. Define $u = 1 + hu_m$ in
the group ring of $C_2\times C_4^m$. Then similar to \sref{products}, $u^2 =
0$, $\rank u = 4^m$, $u$ is symmetric and thus defines a self-dual
code $\C_m$ say.

\begin{theorem}\label{thmref:gf4} $\C_m$ has distance $2^{m+1}$ and is
  thus a $(2\times 4^m, 4^m, 2^{m+1})$ self-dual code.
\end{theorem}
\begin{proof} The proof is very similar to the proof of \thmref{distance} by  
finding the distribution when $m=1$ and then using properties of
direct products. The distribution is slightly better which gives the
better distance. 
\end{proof}

Thus we get $(8, 4, 4), (32, 16, 8), (128, 64, 16)$ etc.\ self-dual
codes over $GF(4)$. These codes use the Euclidean inner product. 
Multiplying the length by 4 multiplies the distance
by 2. 
 
The above examples do not take full advantage of the  symplectic inner
 product. 

For a group ring element $w(\om)$ define $\overline{w} =
w^{-1}(\om^{-1}) = w^{-1}(\om^2)$.  
Define $u_1 = (\om a_1 + \om^2 a_1^3)$ and in general
$u_{i+1} = u_i(\om a_{i+1} + \om^2 a_{i+1}^3)$.  

It is easy to check $u_i^4 = \overline{u_i}^4 =  1$. Define $u=1 +h u_m$
in the group ring of $C_2\ti C_4^m$. Then $u^4 = \overline{u}^4 = 0$, 
$\rank u = \frac{3}{2}\ti 4^m$ and thus $u$ defines a dual-containing code
$\overline{\C_m}$ of rate $\frac{3}{4}$ in the group ring of $C_m^m$
over $GF(4)$ with the symplectic inner product. The proof of the rank
is similar to previous cases.    

\begin{theorem}\label{thmref:gf5} $\overline{\C_m}$ has distance $2^{m}$ and is
  thus a $(2\ti 4^m, \frac{3}{2}\ti 4^m, 2^{m})$ dual-containing  code.
\end{theorem}
\begin{proof} The proof is similar to the proof of \thmref{distance}.
\end{proof}

This gives $(8,6,2), (32,24, 4), (128, 96,8)$ etc.\ dual-containing codes
over $GF(4)$ with the symplectic inner product.

The rates can be extended as follows: Define $u_1 = (\om a_1 + \om^2
a_1^7)$ in the group ring of $C_8$ and in general
$u_{i+1} = u_i(\om a_{i+1} + \om^2 a_{i+1}^7)$ in the group ring of
$C_8^i$. Define $u=1 + h(u_m)$ in the group ring of $C_8^m\ti
C_2$. Then $u^8 = 0$ and $ \rank u = \frac{7}{4}\ti 8^m$. Thus $u$
determines a dual-containing code $\overline{\C_m}$ of rate
$\frac{7}{8}$.

\begin{theorem}\label{thmref:gf6} $\overline{\C_m}$ has distance $2^{m}$ and is
  thus a $(2\ti 8^m, \frac{7}{4}\ti 8^m, 2^{m})$ dual-containing  code.
\end{theorem}

Higher rate dual-containing codes can also be obtained by the methods
of this paper. 





\noindent National University of Ireland, Galway\\Galway\\
Ireland. \\email: ted.hurley@nuigalway.ie
\end{document}